\documentstyle[aps,twocolumn,epsf]{revtex}

\def\be{\begin{equation}}
\def\ee{\end{equation}}

\newcounter{fig}

\catcode`\@=11   

\newcommand{\fcaption}[1]{\vspace{1ex}   
        \refstepcounter{figure}   
        \setbox\@tempboxa = \hbox{\footnotesize {\bf Fig.~\thefigure.} #1}   
        \ifdim \wd\@tempboxa > 8cm   
           {\begin{center}   
        \parbox{8cm}{\footnotesize\baselineskip=8pt {\bf Fig.~\thefigure.} #1}   
            \end{center}}   
        \else   
             {\begin{center}   
             {\footnotesize {\bf Fig.~\thefigure.} #1}   
              \end{center}}   
        \fi}

\begin{document}

\title{Two-Particle Dispersion in Model Velocity Fields
}

\author{I.M.Sokolov
}

\address{Theoretische Polymerphysik, Universit\"{a}t Freiburg, \and %
Hermann-Herder-Str. 3, D-79104 Freiburg i.Br., Germany}

\address{\rm (Received: )}
\address{\mbox{ }}
\address{\parbox{14cm}{\rm \mbox{ }\mbox{ }
We consider two-particle dispersion in a velocity field, where the relative
two-point velocity scales according \ to $v^{2}(r)\propto r^{\alpha }$ and
the corresponding correlation time scales as $\tau (r)\propto r^{\beta }$,
and fix $\alpha =2/3$, as typical for turbulent flows. We show that two
generic types of dispersion behavior arize: For $\alpha /2+\beta <1$ the
correlations in relative velocities decouple and the diffusion approximation
holds. In the opposite case, $\alpha /2+\beta >1,$ the relative motion is
strongly correlated. The case of Kolmogorov flows corresponds to a marginal,
nongeneric situation.
}}
\address{\mbox{ }}
\address{\parbox{14cm}{\rm PACS No:   05.40.-a; 47.27.Qb}}
\maketitle

\makeatletter
\global\@specialpagefalse

\makeatother

\pagebreak

Since the seminal work of Sir L.F.Richardson on particles' dispersion in
atmospheric turbulence [1] a large amount of work has been done in order to
understand the fundamentals of this process (see [2] and [3] for reviews).
Based on empirical evidence, Richardson found out that the mean square
distance $R^{2}(t)=\left\langle r^{2}(t)\right\rangle $ between two
particles dispersed by a turbulent flow grows proportionally to $t^{3}$. The
works of Obukhov and Batchelor have shown that Richardson's law is closely
related to the Kolmogorov-Obukhov scaling of the relative velocities in
turbulent flows. Scaling arguments based on dimensional analysis allow then
to understand the overall type of the behavior of $R^{2}(t)$, but a full
theoretical picture of the dispersion process is still lacking [3].

The theoretical description of dispersion processes typically starts from
models, in which one fixes the spacial statistics of the well developed
turbulent flow (Kolmogorov-Obukhov energy spectrum), and discusses different
types of temporal behavior for the flows [3]. Three situations have been
considered in-detail so far. Here, the white-in time flows represent a toy
model which allows for deep analytical insights [4-6]. In connection with
''real'' turbulence two other cases are widely discussed. One of them
supposes that the temporal decorrelation of the particles' relative motion
happens because the pair as a whole is moving with a mean velocity relative
to an essentially frozen flow structure (as proposed by a Taylor hypothesis)
[7,8]. Another premise connects this decorrelation with the death and birth
of flow structures (''eddies''), whose lifetime is governed by Kolmogorov's
universality assumption [8,9]. Both these situations are extremely awkward
for theoretical analysis.

In the present letter we address the following question: What are the
generic types of two-particle dispersion behavior in a velocity field whose
statistical spatial structure is fixed (and similar to that of a turbulent
flow), if its temporal correlation properties change. This question will be
discussed in the framework of numerical simulations and scaling concepts. As
we proceed to show, two generic types of behavior arise. Thus, the
white-in-time flow and the Taylor-type situation belong to the classes of
diffusive and ballistic behavior, respectively. The case of Kolmogorov
temporal scaling represents a borderline situation.

Let us consider modes of particles' separation in a velocity field whose
two-time correlation function of relative velocities behaves as $%
\left\langle {\bf v}({\bf r},t_{1}) {\bf v}( {\bf r},t_{2})\right\rangle \propto
\left\langle v^{2}(r)\right\rangle g\left[ (t_{2}-t_{1})/\tau (r)\right] $,
where $\tau (r)$ is the distance-dependent correlation time. The $g$%
-function is defined so that $g(0)=1$\ and $\int_{0}^{\infty }g(s)ds=1$. The
mean square relative velocity and the correlation time scale as 
\begin{equation}
\left\langle v^{2}(r)\right\rangle \propto v_{0}^{2}\left( \frac{r}{r_{0}}%
\right) ^{\alpha }  \label{distance}
\end{equation}
and 
\begin{equation}
\tau (r)\propto \tau _{0}\left( \frac{r}{r_{0}}\right) ^{\beta }.
\label{time}
\end{equation}
One can visualize such a flow as being built up from several structures
(plane waves, eddies, etc., see Ref.[3]), each of which is characterized by
its own spatial scale and its scale-dependent correlation time. In
well-developed turbulent flows one has $v^{2}(r)\propto \epsilon
^{2/3}r^{2/3}$, where $\epsilon $ is the energy dissipation rate, so that $%
\alpha =2/3$. The white-in-time flow corresponds to $\beta =0$. Kolmogorov
scaling implies $\beta =2/3$ and Taylor's frozen-flow assumption leads to $%
\beta =1$.

In our simulations we model the two-particle relative motion using the
quasi-Lagrangian approach of Ref.[9]. Parallel to Ref.[9] we confine
ourselves to a two-dimensional case, which is also of high experimental
interest [12, 13]. The relative velocity $\ {\bf v} ( {\bf r}%
,t)=\nabla \times \eta ({\bf r},t) $\ is given by the
quasi-Lagrangian stream function $\eta $. This function is built up from the
contributions of radial octaves: 
\begin{equation}
\eta ({\bf r},t)=\sum\limits_{i=1}^{N}k_{i}^{-(1+\alpha /2)}\eta _{i}(%
k_{i} {\bf r},t),  \label{qlstreem}
\end{equation}
where $k_{i}=2^{i}$, and the flow function for one-octave contribution in
polar coordinates $(r,\theta )$ is given by $\eta _{i}( k_{i} {\bf %
r},t)=F(k_{i}r)\left( A_{i}(t)+B_{i}(t)\cos (2\theta +\phi _{i})\right) $.
The radial part $F(x)$ obeys $F(x)=x^{2}(1-x)$ for $0\leq x\leq 1$ and $%
F(x)=0$ otherwise, and $\phi _{i}$ are quenched random phases. Moreover, $%
A_{i}(t)\,$and $B_{i}(t)$ are independent Gaussian random processes with
dispersions $A^{2}=B^{2}=v_{0}^{2}$ and with correlation times $\tau
_{i}=2^{-i\beta }\tau _{0}$. At each time step these processes are generated
according to $X_{i}(t+\Delta t)=\sqrt{1-(\Delta t/\tau _{i})^{2}}%
X_{i}(t)+(\Delta t/\tau _{i})v_{0}^{2}\zeta $, where $X$ is $A$ or $B,$ and $%
\zeta $ is a Gaussian random variable with zero mean and unit variance. The
values of $\tau _{0}$ and the integration step $\Delta t$ are to be chosen
in such a way that $\tau _{N}\gtrsim \Delta t$. Typically, values of $\Delta
t\sim 10^{-4}$ are used. For the noncorrelated flow ($\beta =0$) the values
of $A_{i}(t)\,$and $B_{i}(t)$ are renewed at each integration step $\Delta t$%
. In the present simulations $N=16$ was used. The value $v_{0}=1$ was
employed in the majority of simulations reported here, so that only the use
of a different $v_{0}$ value will be is explicitly stated in the following.

The values of $R^{2}(t)$ obtained from 3000 realizations of the flow for
several values of $\beta \in [0,1]$ are plotted on double logarithmic scales
in Fig.1, where $\tau _{0}=0.15$ is used. One can clearly see that for all $%
\beta $ a scaling regime $R^{2}(t)\propto t^{\gamma }$ appears. We note
moreover that the curves for $\beta =0.67,0.8,09$ and $1.0$ are almost
indistinguishable within statistical errors. The values of $\gamma $ as a
function of $\beta $ are presented in the insert, together with the
theoretically predicted forms, vide infra.

\begin{figure} \begin{center}   
  \fbox{\epsfysize=8cm\epsfbox{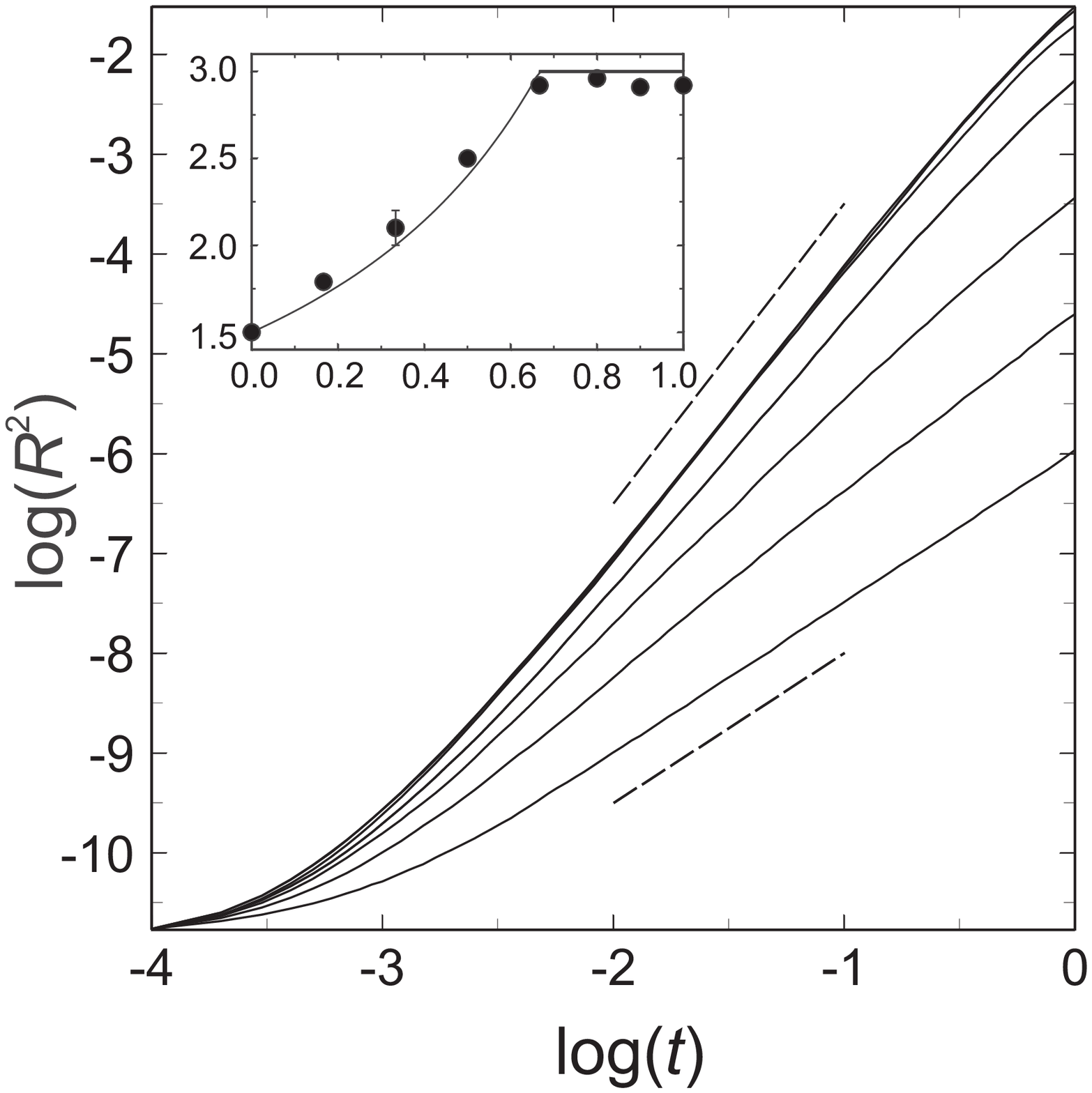}}   
   \fcaption{Mean square displacement $R^{2}(t)$ plotted in double logarithmic
scales. The four lower curves correspond to $\beta =0$, 0.17, 0.33 and 0.5
(from bottom to top). The dashed lines indicate the slopes 1.5 and 3. The
four upper curves for $\beta =0.67$, 0.8, 0.9 and 1 are hardly
distinguishable within the statistical errors of the simulations. The insert
shows the values of $\gamma (\beta )$ determined by a least-squares-fit
within the scaling region of each curve. The error bar shows the typical
accuracy of all $\gamma $-values. The full lines give the theoretical
predictions, Eq.(\ref{Diffreg}) and Eq.(\ref{Ballistic}).}   
  \label{tra}   
\end{center}\end{figure}

The regimes of dispersion found in simulations can be explained within the
framework put forward in Ref.[10]. The discussion starts by considering $%
l(r)=v(r)\tau (r),$ the mean free path of motion at the distance $r$. If
this mean free path always stays small compared to $r$, the relative motion
exhibits a diffusive behavior with a position-dependent diffusion
coefficient, $K(r)\propto l^{2}(r)/\tau (r)\propto r^{\alpha +\beta }$.
Taking as a scaling assumption $r\propto \left\langle r^{2}(t)\right\rangle
^{1/2}=R$, one gets that the mean square\ separation $R$ grows as $%
R^{2}\propto t^{\gamma }$ with 
\begin{equation}
\gamma =\frac{2}{2-(\alpha +\beta )}.  \label{Diffreg}
\end{equation}
On the other hand, if $l(r)$ is of the order of $r$, the mean separation
follows from the integration of the ballistic equation of motion $\frac{d}{dt%
}R=v(R)\propto R^{\alpha /2}$, see Ref.[11]. Thus, in a flow where a
considerable amount of flow lines of relative velocity are open, one gets $%
R^{2}\propto t^{\gamma },$ with 
\begin{equation}
\gamma =\frac{4}{2-\alpha }.  \label{Ballistic}
\end{equation}
The occurrence of either regime is governed by the value of the (local)
persistence parameter of the flow, 
\begin{equation}
Ps(r)=l(r)/r=v(r)\tau (r)/r.
\end{equation}
Small values of $Ps$ correspond to erratic, diffusive motion, while large
values of $Ps$ imply that the motion is strongly persistent. The value of
the persistence parameter scales with $r$ as $Ps(r)\propto r^{\alpha
/2+\beta -1}$. Since under particle's dispersion the mean interparticle
distance grows continuously with time, the value of $Ps$ decreases
continuously for $\alpha /2+\beta <1$, so that the diffusive approximation
is asymptotically exact. For $\alpha /2+\beta >1$ the lifetimes of the
structures grow so fast that the diffusive approximation does not hold. This
situation is one observed in our simulations for $\beta >2/3$. The strong
ballistic component of motion implies that the velocities stay correlated
over considerable time intervals. The results of Fig.1 confirm that $\gamma
(\beta )$ behaves accordingly to Eq.(\ref{Diffreg}) for $\beta <2/3$ and Eq.(%
\ref{Ballistic}) for $\beta >2/3$. We note here that the parameters of the
simulations presented in Fig.1 ($v_{0}=1,$ $\tau _{0}=0.15$) were chosen in
a way that allows to show all curves within the same time- and distance
intervals. This leads to a somehow restricted scaling range and to slight
overestimate of $\gamma $-values in the diffusive domain.

Strong differences between the diffusive and the ballistic regimes can be
readily inferred when looking at typical trajectories of the motion, such as
are plotted in Fig.2 for the cases $\beta =0.33$ and $\beta =0.67$. The
difference between the trajectories is evident both in the $(x,y)$-plots and
in the $r(t)$-dependences. The curves for $\beta =0.33$ exhibit a
random-walk-like, erratic behavior, while the curves for $\beta =0.67$ show
long periods of laminar, directed motion. In order to quantitatively
characterize the strength of the velocity correlations we calculate the
backwards-in-time correlation function (BCF) of the radial 
velocities, as introduced in Ref.[12]. This function is defined as $%
C_{r}(\tau )=\left\langle v_{r}(t-\tau )v_{r}(t)\right\rangle /\left\langle
v_{r}^{2}(t)\right\rangle $ and shows, what part of its history is
remembered by a particle in motion. The function is plotted in Fig.3 against
the dimensionless parameter $\vartheta =-\tau /t$. The functions (obtained
in $10^{4}$ realizations each) are plotted for 4 different sets of
parameters. Here the dashed lines correspond to $\beta =0.33$, in the
diffusive range, for $t=10^{-2},$ $3\cdot 10^{-2},$ $10^{-1}$ and $3\cdot
10^{-1}$. These BCF do not scale and are rather sharply peaked close to
zero, thus indicating the loss of memory. The two sets of full lines
indicate $C_{r}(\tau )$ in Kolmogorov flows, for $t=10^{-2},$ $3\cdot
10^{-2},$ and $10^{-1}$. The lower set corresponds to the value $\tau
_{0}=0.05$ and the upper set to the value $\tau _{0}=0.15$. In both cases
the functions show scaling behavior. No considerable changes in the BCF's
form occur when further increasing the value of $\tau _{0}$ up to $\tau
_{0}=1$, thus indicating that the data $\tau _{0}=0.15$ correspond already
to a strongly correlated regime. The form of these curves resembles closely
the experimental findings of Ref. [12]. The BCFs for $\beta =1.0\,$show an
overall behavior very similar to the one in Kolmogorov's case. Note that as
the time grows the curves for $\beta =1.0$ approach those for $\beta =2/3$
and probably tend to the same limit. The curves for $\beta =1.0$ and $\tau
_{0}=1$ (not shown) fall together with those in Kolmogorov's case with $\tau
_{0}=0.15$.

\begin{figure} \begin{center}   
  \fbox{\epsfysize=6cm\epsfbox{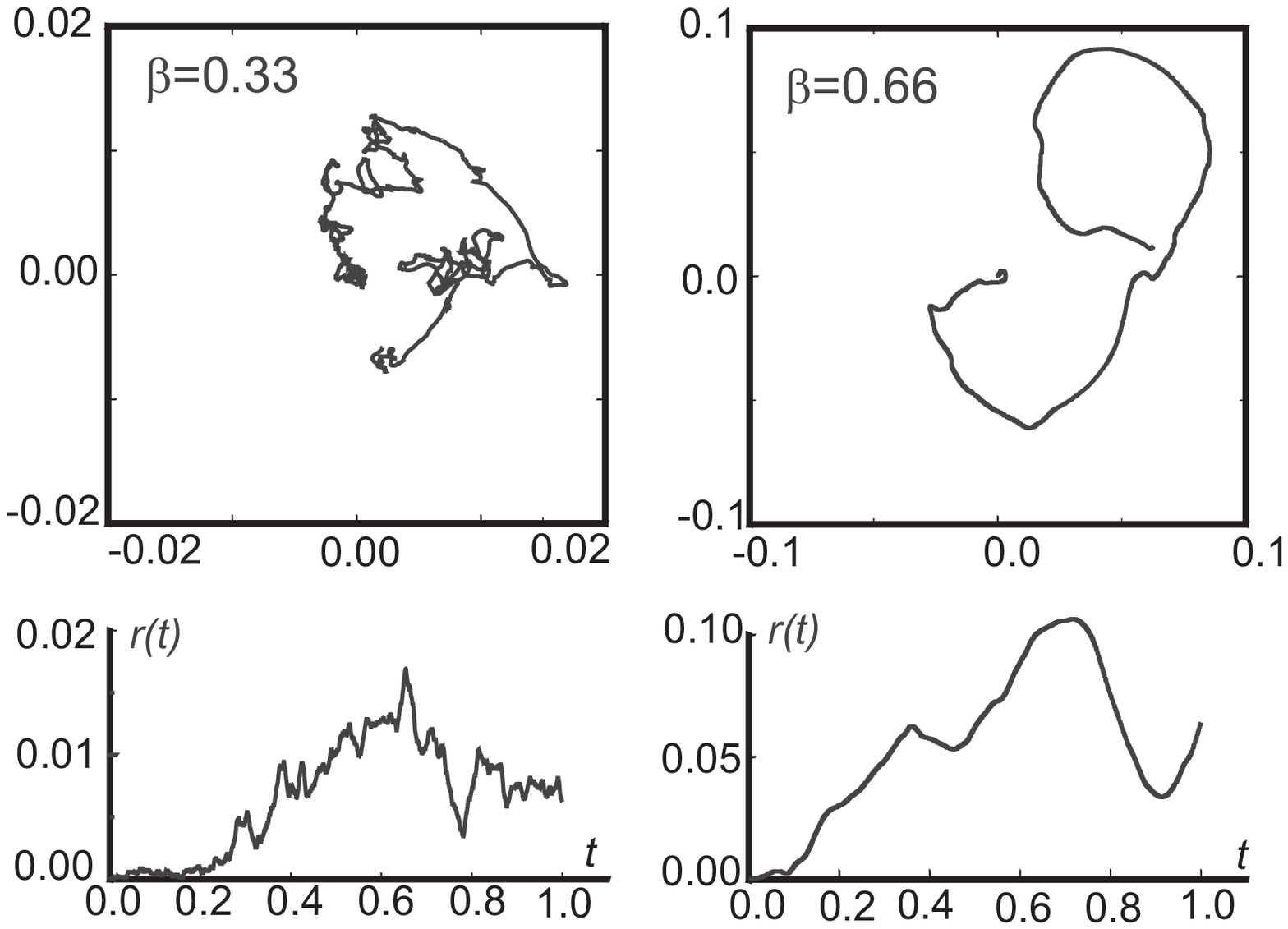}}   
   \fcaption{Typical trajectories in the diffusive regime ($\beta =0.33$) and in
the Kolmogorov regime ($\beta =0.67$). The upper pictures show the
trajectories in the $(x,y)$-plane, the lower ones represent the
corresponding $r(t)$-behavior. Note that the scales of the right and of the
left graphs differ by a factor of 5.}   
\label{tri}   
\end{center}\end{figure}

The similarity in the properties of dispersion processes in a Kolmogorov
situation with larger $Ps$ (larger $\tau _{0}$) and in ballistic regime can
be explained based on the behavior of the effective persistence parameter.
In the diffusive regime we supposed that the correlation time of the
particles' relative velocity scales in the same way as the Eulerian lifetime
of the corresponding structures. On the other hand, in the ballistic regime, 
$\beta >1-\alpha /2$, the lifetimes of the structures grow so fast that no
considerable decorrelation takes place during the time the particles sweep
through the structure. The Lagrangian decorrelation process is then
connected not to Eulerian decorrelation, but to sweeping along open flow
lines. The effective correlation time then scales according to $\tau
_{s}(r)\propto r/v(r)\propto t^{1-\alpha /2}$, and the effective value of $%
\beta $ stagnates at $\beta =1-\alpha /2$. Thus, all long-time correlated
cases belong to the same universality class of strongly-correlated flows, as
the Kolmogorov flows with large $Ps$, for which Eq.(\ref{Diffreg}) and (\ref
{Ballistic}) coincide.

\begin{figure} \begin{center}   
  \fbox{\epsfysize=6cm\epsfbox{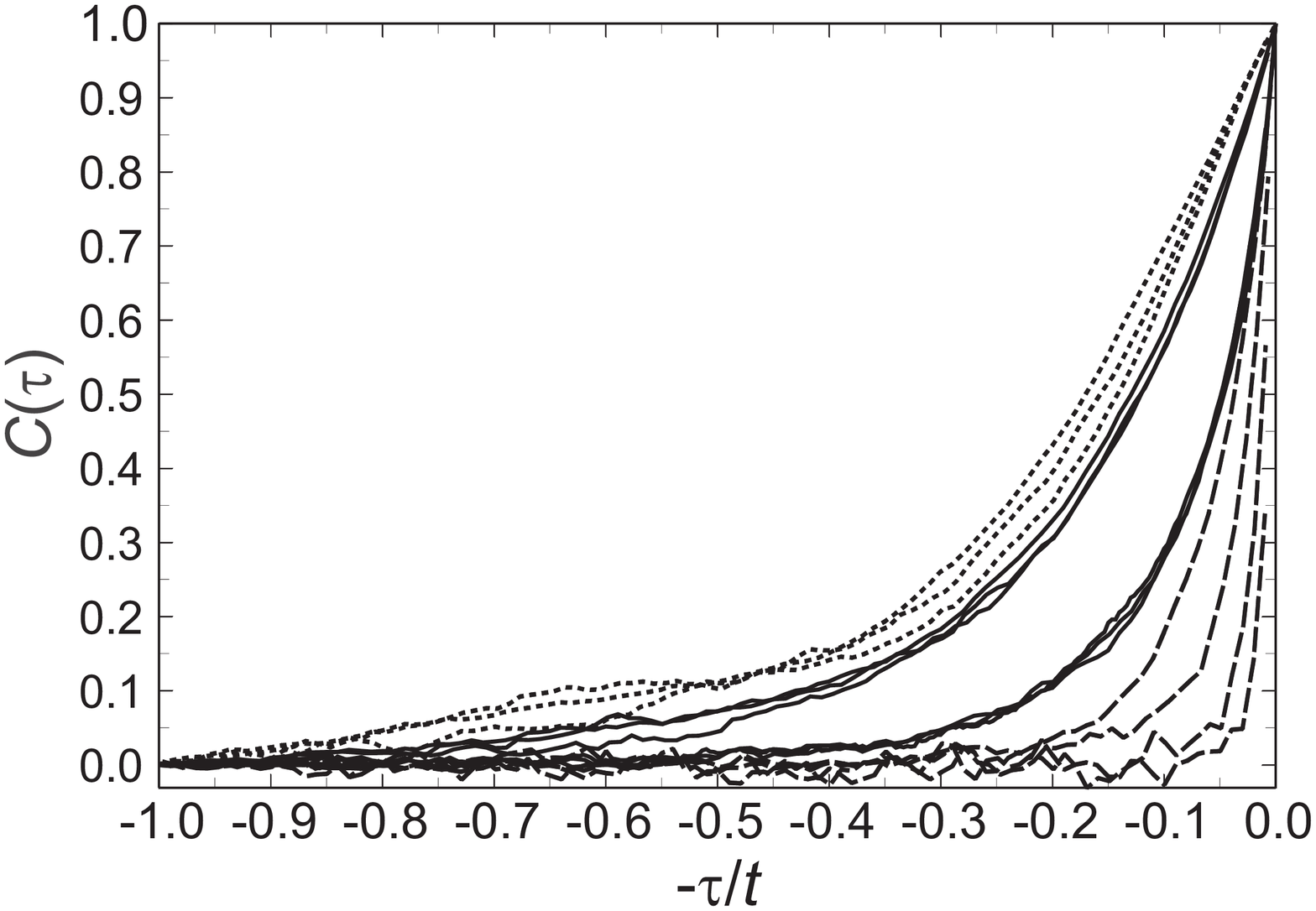}}   
   \fcaption{The BCF of relative velocities as a function of the dimensionless
time lag $-\tau /t$. The lower group of dashed lines corresponds to $\beta
=0.33$ (the values of $t$ are 0.01, 0.03, 0.1 and 0.3, from top to bottom).
The two groups of full curves corresponds to the Kolmogorov case (three
curves for $t=0.01, 0.03, 0.1 $ each). The dotted curves correspond to $%
\beta =1$ for the same values of time, see text for details.}   
  \label{tru}   
\end{center}\end{figure}

For Kolmogorov flows the ballistic and the diffusive mechanisms lead to the
same functional form of $R^{2}(t)$-dependence. The functional form of the
dependence of $R^{2}(t)$ on parameters of the flow is $R^{2}\propto \left(
v_{0}^{2}\tau _{0}/r_{0}^{\alpha +\beta }\right) ^{\gamma }t^{\gamma }$ in
the diffusive situation ($Ps\ll 1$) and $R^{2}\propto \left(
v_{0}/r_{0}^{\alpha /2}\right) ^{\gamma }t^{\gamma }$ in the ballistic case (%
$Ps\gg 1$). Assuming that $Ps$ is the single relevant parameter governing
the dispersion we are lead to the form $R^{2}(t)\propto f(Ps)\left(
v_{0}/r_{0}^{\alpha /2}\right) ^{\gamma }t^{\gamma }$, where $f(Ps)$ is a
universal function of $Ps$, which behaves as $Ps^{\gamma }$ for $Ps\ll 1$
and tends to a constant for $Ps\gg 1$. Thus, for a fixed spatial structure of
the flow, the following scaling assumption is supposed to hold: 
\begin{equation}
\frac{R^{2}(t)}{\left( v_{0}t\right) ^{\gamma }}=F(v_{0}\tau _{0}),
\label{scaling}
\end{equation}
which scaling can be checked in our case by plotting $R^{2}(t)/(v_{0}t)^{3}$
against $v_{0}\tau _{0}$. The corresponding plot is given in Fig. 4, where
we fix $t=0.1$, and plot the results in three series of simulations. Each
point corresponds to an average over $5\cdot 10^{4}$ runs. Here the squares
correspond to $v_{0}=1$ and to the values of $\tau _{0}$ ranging between
0.01 and 0.15, the triangles correspond to $v_{0}=0.3$ and th $\tau _{0}$
between 0.033 and 0.5, and the circles to $\tau_{0}=0.1$ and to values of $
v_{0}$ between 0.1 and 1.5. The error bar indicates a typical statistical
error as inferred from 5 similar series of $5\cdot 10^{4}$ runs each. The
scaling proposed by Eq.(\ref{scaling}) is well-obeyed by the results. Some
points outside of the range of Fig.4 were also checked. Thus, for larger
values of $v_{0}\tau _{0}$ the values of $R^{2}(t)/(v_{0}t)^{3}$ seem to
stagnate. On the other hand, increasing $v_{0}\tau _{0}$ to values larger
than 0.3 (i.e. approaching the frozen flow regime) leads to a strong
increase in fluctuations, making the results less reliable.

\begin{figure} \begin{center}   
  \fbox{\epsfysize=6cm\epsfbox{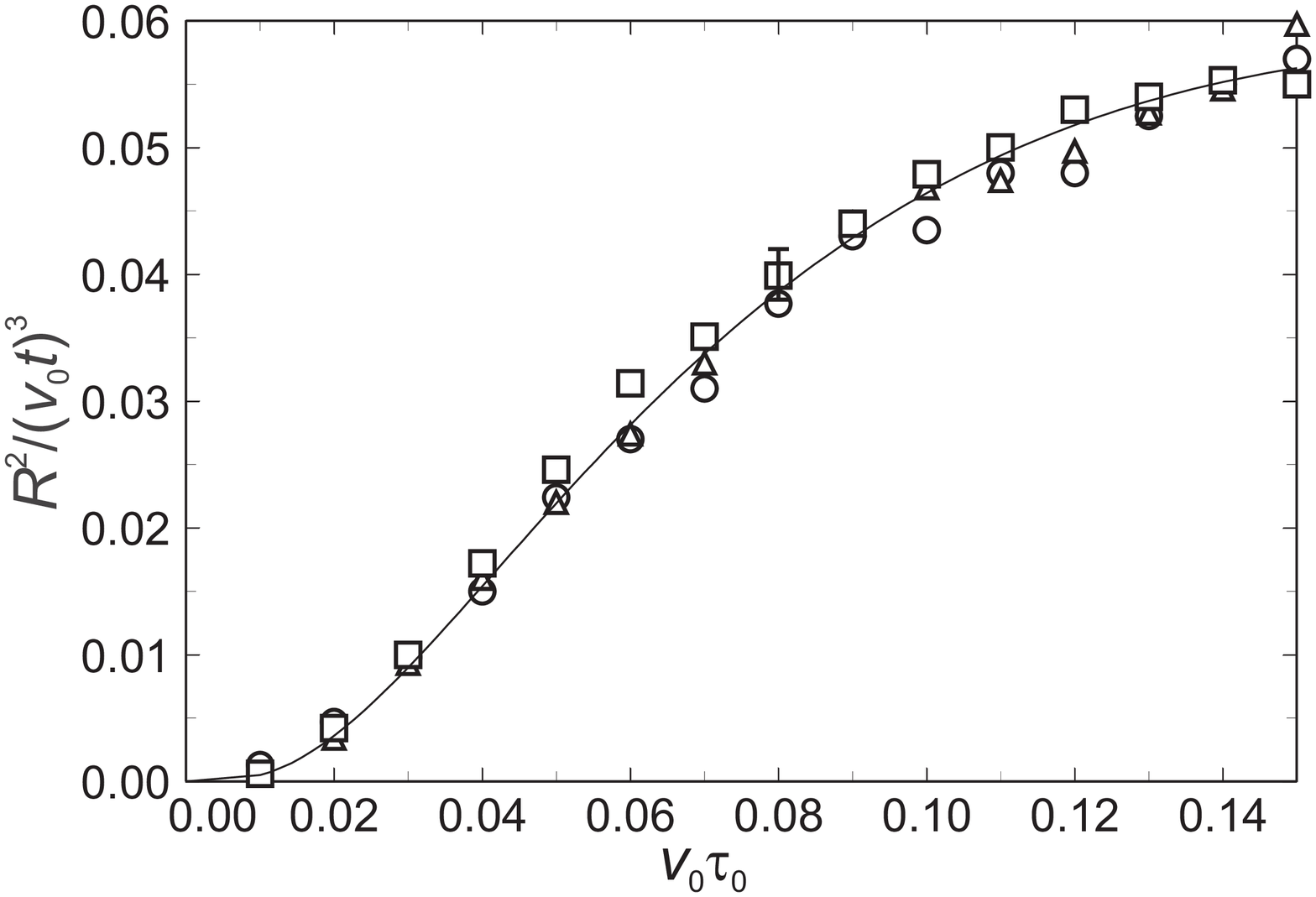}}   
   \fcaption{The values of $R^{2}(t)/(v_{0}t)^{3}$ plotted against $v_{0}\tau _{0}$. 
The dashed line is drawn as a guide to the eye.}   
  \label{trau}   
\end{center}\end{figure}

Let us summarize our findings. Thus, we considered two-particle dispersion
in a velocity field scaling according to $v^{2}(r)\propto r^{2/3}$ and $\tau
(r)\propto r^{\beta }$. We show that two generic types of behavior are
possible: For $\alpha /2+\beta <1$ the diffusion approximation holds and the
increase in the interparticle distances is governed by the
distance-dependent diffusion coefficient $K(r)\propto r^{\alpha +\beta }$.
In the opposite case $\alpha /2+\beta >1$ the relative velocities stay
strongly correlated. The transition between the two regimes takes place
exactly for the Kolmogorov flow, for which $\alpha /2+\beta =1$. In this
case the properties of the dispersion process depend on the persistence
parameter of the flow.

The author is thankful to P.Tabeling, I.Procaccia, V.L'vov, J.Klafter and A.
Blumen for enlightening discussions. Financial support by the Deutsche
Forschungsgemeinschaft through the SFB 428 and by the Fonds der Chemischen
Industrie is gratefully acknowledged.

\end{document}